\documentclass[%
 reprint,
preprintnumbers,
nofootinbib,
 amsmath,amssymb,
 aps,
]{revtex4-2}

\usepackage{graphicx}
\usepackage{dcolumn}
\usepackage{bm}
\usepackage{multirow}
\usepackage{hyperref}
\usepackage{subfig}
\usepackage{feynmp-auto}
\usepackage{cleveref}

\begin{document}

\preprint{CP3-19-38}
\preprint{MCnet-19-19}
\preprint{ICPP-016}
\preprint{IPPP/19/67}

\title{Towards a generic implementation of matrix-element maximisation as a classifier in particle physics}

\author{Stefan von Buddenbrock}
\email{stef.von.b@cern.ch}
\affiliation{School of Physics and Institute for Collider Particle Physics, University of the Witwatersrand, Johannesburg, Wits 2050, South Africa}
\affiliation{Centre for Cosmology, Particle Physics and Phenomenology (CP3/IRMP) Universit\'e Catholique de Louvain, 1348 Louvain-la-neuve, Belgium}

\author{Olivier Mattelaer}
\email{olivier.mattelaer@uclouvain.be}
\affiliation{Centre for Cosmology, Particle Physics and Phenomenology (CP3/IRMP) Universit\'e Catholique de Louvain, 1348 Louvain-la-neuve, Belgium}

\author{Michael Spannowsky}
\email{michael.spannowsky@durham.ac.uk}
\affiliation{Institute for Particle Physics Phenomenology, Department of Physics, Durham University, Durham DH1 3LE, U.K.}

\date{\today}

\begin{abstract}
The so-called matrix-element method (MEM) has long been used successfully as a classification tool in particle physics searches.
In the presence of invisible final state particles, the traditional MEM typically assigns probabilities to an event -- based on whether it is more signal or background-like -- through a phase space integration over all degrees of freedom of the invisible particles in the process(es).
One inherent shortcoming of the traditional MEM is that the phase space integration can be slow, and therefore impractical for high multiplicity final states and/or large data sets.
The recent alternative of matrix-element maximisation has recently been introduced to circumvent this problem, since maximising a highly-dimensional function can be a far more CPU-efficient task than that of integration.
In this work, matrix-element maximisation is applied to the process of fully-leptonic top associated Higgs production, where the Higgs boson decays to two $b$-quarks.
A variety of optimisation algorithms are tested in terms of their performance and speed, and it is explicitly found that the maximisation technique is far more CPU-efficient than the traditional MEM at the cost of a slight reduction in performance.
An interesting consequence of using matrix-element maximisation is that the result of the procedure gives an estimate of the four-momenta for the invisible particles in the event.
As a result, the idea of using these estimates as input information for more complicated tools is discussed with potential prospects for future developments of the method.

\end{abstract}


\maketitle


\section{Introduction}

The detailed analysis of measured data, with the aim to find new physics disguised as novel resonances or coupling-deviations of known particles, is of crucial importance for experimental collaborations at the LHC and theorists alike. 
Novel strategies have been proposed to compare final states according to their probability of being produced from competing hypotheses, e.g. the Standard Model hypothesis with the hypothesis of a new physics model. 
Two of the most promising strategies that have transpired over recent years to perform an automated and sensitive way of performing this task are matrix-element based methods \cite{Kondo:1988yd,Abazov:2004cs,Artoisenet:2010cn} and machine-learning based methods \cite{Albertsson:2018maf,Brehmer:2019xox}.

Matrix-Element Methods (MEMs) rely on first-principle calculated matrix-elements for signal and background hypotheses. 
As such the MEM has no requirement for training on pseudo-data generated by event generators, but can be directly applied to measured data. 
The measured final state objects' momenta for the processes of interest are then used as an input to the matrix-elements. 
This allows one to construct an ideal classifier \cite{James:2000et}, based on the likelihood ratio between signal and background. 
Thus, the MEM has been used very successfully in a wider range of applications and measurements \cite{Cranmer:2006zs,Alwall:2010cq,Alwall:2009sv,Andersen:2012kn,Artoisenet:2013vfa,Betancur:2017kqe}, and has recently been extended to the substructure of jets \cite{Soper:2011cr,Soper:2012pb,deLima:2014dta,FerreiradeLima:2016gcz}, next-to-leading order accuracy \cite{Campbell:2012cz,Campbell:2013hz,Martini:2015fsa,Gritsan:2016hjl,Martini:2017ydu,Kraus:2019qoq} and even to fully exclusive final states with an arbitrary number of reconstructed objects \cite{Soper:2014rya, Englert:2015dlp}. 

An inherent shortcoming of the MEM is the time it takes to evaluate its weights for high-multiplicity final states. 
As the measured momenta of the final states $K_i$ differ from the momenta associated with the hard interaction of the process $k_i$ (due to effects from soft/co-linear emissions, hadronisation and experimental reconstruction deficiencies) one needs to introduce transfer functions $W(k_i,K_i)$ to account for this difference. 
The difference between $k_i$ and $K_i$ is more pronounced for jets than for isolated photons and leptons. 
These often Gaussian-shaped transfer functions are then convoluted with the squared matrix-elements of the hard interaction to calculate the probability of the measured final state being produced by a certain signal or background hypothesis. 
This convolution is performed via Monte-Carlo integration over three dimensions, e.g. $p_\text{T}$, $\phi$ and $y$, for each measured final state object, which can amount to a highly non-trivial and computation-time expensive task. 
When multiple invisible particles are a result of the hard interaction, this problem is significantly amplified as the whole matrix-element (including the convolution with the transfer functions) has to be integrated over the entire allowed phase space available to the invisible particles. 
Thus, two major obstacles in the application of MEMs in the classification of complex final states are the computationally expensive tasks of (i) including transfer functions and (ii) integrating over the phase space of invisible particles. 
Thus, the applicability of the MEM in searches and measurements depends crucially on whether it can be automated and applied to complex final states. 

MEMs like Shower/Event Deconstruction~\cite{Soper:2011cr,Soper:2012pb,Soper:2014rya,Englert:2015dlp} and their extension to multi-jet merged matrix-elements, as in HYTREES \cite{Prestel:2019neg}, can be applied to fully showered and hadronised final states. 
Such methods access more information and should result in an improved classification performance compared to the fixed-order MEMs. 
However, the complexity of such final states render the use of transfer functions prohibitive. 
Fortunately, after calculating the matrix-elements including the perturbative parton shower weights down to the hadronisation scale of $\mathcal{O}(10)$ GeV,  transfer functions have a severely diminished effect on the kinematics, i.e. here $k_i \simeq K_i$, and it suffices to replace the physical width of decaying resonances by their experimentally measured width to account for an imperfect experimental reconstruction in phase space regions where the matrix-element varies rapidly.
However, as the likelihood ratios for signal and background are evaluated by summing over all semi-classically allowed paths that connect the hypothesised initial states with the measured final states, quickly millions of trees have to be calculated. 
For these methods an integration over the whole phase space of invisible particles while summing over all possible trees is a serious technical challenge.
We therefore propose to avoid such an integration and define the momenta of the invisible particles early on in the calculation by a likelihood maximisation procedure \cite{FerreiradeLima:2017iwx}. 
After determining these momenta in this way they can be included in the calculation of the probability trees straightforwardly.

In \Cref{sec:method} we will describe the method and apply it to the phenomenologically relevant process of $t\bar{t} (h \to b\bar{b})$ production with leptonic top quark decays in \Cref{sec:tth}.
We will compare our method with the traditional MEM for this process and discuss advantages and disadvantages of each approach. 
The article is concluded with a summary and some future prospects in \Cref{sec:summary}.

\section{Methodology}
\label{sec:method}

Following on from the method introduced in Ref.~\cite{FerreiradeLima:2017iwx}, this work makes use of the MEM in terms of maximisation as opposed to phase space integration.
For a given scattering process $\alpha$, the matrix-element $\mathcal{M}_\alpha$ is a complex-valued functional defined in terms the initial and final state momenta of all the particles involved.
The mod-square of $\mathcal{M}_\alpha$ is related to the probability density of the process occurring in a given region of the phase space.
In this work, we use squared matrix-elements as computed by \texttt{MadGraph5\_aMC@NLO}~\cite{Alwall:2014hca} (dubbed \texttt{MG5aMC}).

Given an input phase space point $x$, a weight is assigned to it by maximising an objective function over the Lorentz invariant phase space $\Phi$ in the following way:
\begin{equation}
    w_\alpha(x)=\max_{y\in \Phi} \left\{\left|\mathcal{M}_\alpha\right|^2(y)W(x,y)\right\}.
    \label{eqn:maximisation}
\end{equation}
The transfer function $W(x,y)$ encodes information about the resolution of the input event $x$.
Technically speaking, it is the probability that a test phase space point $y$ is a fluctuation of $x$ due to experimental detector resolution.
Note that the input phase space point $x$ is assumed to be composed of experimental information, and therefore do not contain any information about invisible particles that might be produced.
The result of the maximisation process provides an estimate of the four-momentum for each invisible particle in the event (this will be discussed in\Cref{sec:tth}).

The maximisation procedure in \Cref{eqn:maximisation} is most efficient if the objective function has a structure that peaks in its maximisation variables.
The transfer function $W$ is (roughly speaking) the product of Gaussian-like probability density functions, and is therefore relatively simple to maximise.
On the other hand, the squared matrix-element $\left|\mathcal{M}\right|^2$ does not necessarily peak in its standard maximisation variables (that is, the four-momenta of the initial and final state particles) and is therefore not an efficient function to maximise.
If one performs a change of variables, the structure of the squared matrix-element can be transformed to something that can be more efficiently maximised.
The obvious way of doing this is to exploit the Breit-Wigner mass peaks for each resonance defined in the process, as done in the \textit{block} formalism in \texttt{MadWeight}~\cite{Artoisenet:2010cn}.
In short, the initial and final state momenta can be mapped onto a set of coordinates that include all of the resonance masses related to the propagators in the relevant Feynman diagrams, thereby giving the squared matrix-element a highly peaked structure.

By performing such a change of variables and encoding the appropriate transfer function, the efficiency of the maximisation process can be made even more efficient by mapping each coordinate to the range $(0,1)$.
Then the phase space region is approximately evenly distributed in a multi-dimensional cube of unit length.
This is done by constructing a model of coordinates (i.e. Breit-Wigner peaks for the resonance masses and Gaussians for the transfer function), and then computing the cumulative distribution function for each component of the model.
Assuming that the probability density functions used in the model are normalised, the cumulative density functions will uniquely map numbers in the range $(0,1)$ to the coordinate values distributed by the probability density function in question.

Finding the global maximum of the matrix-element over the multi-dimensional phase space is an optimisation task that can be performed algorithmically in various ways. 
According to the ``no-free-lunch theorem'' of optimisation~\cite{nofreelunch}, there should be no \textit{a priori} choice of an optimal maximisation algorithm, and so a large number of different algorithms were tested. We compared their performance with respect to reconstruction efficiency and speed. 
The \texttt{NLOPT} library of optimisation algorithms was used~\cite{nlopt}, which provides a wide variety of different types of optimisation algorithms.
Since \texttt{MG5aMC} does not provide a derivative function for the squared matrix-element,\footnote{In principle, the derivative could be calculated and used, which would speed up the maximisation process greatly. This may be explored in future.} only the derivative-free algorithms were tested.
A list of all the derivative-free optimisation algorithms from the \texttt{NLOPT} library considered in this work is shown in \Cref{tab:nlopt_algorithms}.
These are divided up into \textit{local} and \textit{global} algorithms.
The former use information about test points and their near neighbours (usually starting from an initial guess), and are therefore more susceptible to finding local maxima.
The latter are designed to probe the maximisation region more evenly and are therefore better at finding global maxima, however at the cost of more computing time.

\begin{table}
    \centering
    \begin{tabular}{lr}
        \textbf{\texttt{NLOPT} identifier} & \textbf{Ref.}  \\
        \hline
        \multicolumn{2}{c}{\it Global maximisation algorithms} \\
        \hline
        \texttt{GN\_CRS2\_LM} & \cite{10.1093/comjnl/20.4.367,Price1983,Kaelo2006} \\
        \texttt{GN\_DIRECT} & \cite{Jones1993}  \\
        \texttt{GN\_DIRECT\_L} & \cite{Gablonsky2001} \\
        \texttt{GN\_DIRECT\_L\_NOSCAL} & \cite{Gablonsky2001} \\
        \texttt{GN\_DIRECT\_L\_RAND} & \cite{Gablonsky2001} \\
        \texttt{GN\_DIRECT\_L\_RAND\_NOSCAL} & \cite{Gablonsky2001} \\
        \texttt{GN\_DIRECT\_NOSCAL} & \cite{Jones1993} \\
        \texttt{GN\_ORIG\_DIRECT} & \cite{Jones1993} \\
        \texttt{GN\_ORIG\_DIRECT\_L} & \cite{Gablonsky2001} \\
        \texttt{GN\_MLSL} & \cite{RinnooyKan1987.1,RinnooyKan1987.2} \\
        \texttt{GN\_MLSL\_LDS} & \cite{Kucherenko2005} \\
        \texttt{GN\_ISRES} & \cite{873238,1424197} \\
        \texttt{GN\_ESCH} & \cite{Beyer2002,5482078} \\
        \hline
        \multicolumn{2}{c}{\it Local maximisation algorithms} \\
        \hline
        \texttt{LN\_AUGLAG} & \cite{doi:10.1137/0728030,doi:10.1080/10556780701577730} \\
        \texttt{LN\_AUGLAG\_EQ} & \cite{doi:10.1137/0728030,doi:10.1080/10556780701577730} \\
        \texttt{LN\_BOBYQA} & \cite{Powell2009TheBA} \\
        \texttt{LN\_COBYLA} & \cite{Powell1994} \\
        \texttt{LN\_NELDERMEAD} & \cite{10.1093/comjnl/7.4.308,10.1093/comjnl/8.1.42} \\
        \texttt{LN\_NEWUOA} & \cite{Powell2006} \\
        \texttt{LN\_NEWUOA\_BOUND} & \cite{Powell2006} \\
        \texttt{LN\_PRAXIS} & \cite{praxis} \\
        \texttt{LN\_SBPLX} & \cite{Rowan90functionalstability} \\
        \hline
    \end{tabular}
    \caption{A list of the derivative-free optimisation algorithms provided in the \texttt{NLOPT} library.}
    \label{tab:nlopt_algorithms}
\end{table}

The weight computed in \Cref{eqn:maximisation} can be thought of as an un-normalised probability.
In order to calculate the formal probability, one should divide the weight by an appropriately calculated visible cross section $\sigma_\alpha^\text{vis}$ (see the definition in Ref.~\cite{Artoisenet:2010cn}).
In this work, we do not perform this normalisation since it is still possible to construct a hypothesis classifier without it.
Given the signal and background hypotheses ($\alpha=s$ and $b$, respectively), the weights $w_s$ and $w_b$ can be calculated using \Cref{eqn:maximisation} for an arbitrary number of input events.
For each event, a simplified discrimination variable $\chi$ can be calculated as follows:
\begin{equation}
    \chi(x)=\frac{\log{w_s(x)}}{\log{w_b(x)}}.
    \label{eqn:chi_discriminant}
\end{equation}
Thereafter, events that are larger in $\chi$ contain phase space information that is more signal-like and events lower in $\chi$ more background-like.
One can then determine a lower cut value $\chi_0$ that maximises a discovery significance for the signal hypothesis (given the correctly calculated cross sections and efficiencies).
This cut value essentially acts as a separation line between signal-like events and background-like events.
Note that this is a simplified one-dimensional classifier; the classification can be further improved by incorporating higher dimensional separation criteria.

\section{Application to the top associated Higgs production process}
\label{sec:tth}

In order to fully test the proposed matrix-element maximisation procedure, we consider the production of the Higgs boson in association with a pair of top quarks ($tth$) at the Large Hadron Collider (LHC).
Furthermore, the Higgs boson $h$ is considered in the $h\to b\bar{b}$ channel, and both top quarks decay leptonically. 
The exclusive final state is therefore quite complicated, consisting of 2 neutrinos, 2 leptons and 4 $b$-tagged jets. This process if of paramount importance to simultaneously test the Yukawa coupling of top and bottom quarks.
This final state has been explored using the traditional MEM by the ATLAS~\cite{Aaboud:2017rss} and CMS~\cite{Sirunyan:2018mvw} experiments, the former having noted that the full functionality of the traditional MEM was limited by computing resources.
In this section, it shall be shown that making use of matrix-element maximisation can provide a comparable result to the traditional MEM with a drastically improved speed.
In addition to this, some salient points about the physical implications of the maximisation procedure are discussed.

\subsection{The control set-up}
\label{sec:control}

\begin{figure*}
    \centering
    \subfloat[Signal process.]{
    \begin{fmffile}{tth}
    \fmfframe(0,10)(0,17){
    \begin{fmfgraph*}(150,100)
    \fmfleft{space01,i2,i1,space02}
    \fmfright{p5,p3}
    \fmftop{space11,space12,space13,p4,p1}
    \fmfbottom{space21,space22,space23,p6,p2}
    \fmf{gluon}{gg1,i1}
    \fmf{gluon}{i2,gg1}
    \fmf{gluon,tension=2}{gg2,gg1}
    \fmf{fermion,label=$t$,l.side=left,tension=2}{gg2,s134}
    \fmf{fermion,label=$\bar{t}$,l.side=left,tension=2}{s256,gg2}
    \fmf{boson,label=$W^+$,l.side=right,tension=1.5,l.dist=1pt}{s134,s13}
    \fmflabel{$b$}{p4}
    \fmf{fermion}{s134,p4}
    \fmf{boson,label=$W^-$,l.side=left,tension=1.5,l.dist=2pt}{s256,s25}
    \fmflabel{$\bar{b}$}{p6}
    \fmf{fermion}{p6,s256}
    \fmflabel{$\ell^+$}{p3}
    \fmf{fermion}{p3,s13}
    \fmflabel{$\ell^-$}{p5}
    \fmf{fermion}{s25,p5}
    \fmflabel{$\nu$}{p1}
    \fmf{fermion}{s13,p1}
    \fmflabel{$\bar{\nu}$}{p2}
    \fmf{fermion}{p2,s25}
    \fmffreeze
    \fmfright{tth1,tth2,gb1,gb2}
    \fmfforce{(0.61w,0.58h)}{tth1}
    \fmfforce{(0.8w,0.58h)}{tth2}
    \fmfforce{(0.95w,0.64h)}{gb1}
    \fmfforce{(0.95w,0.52h)}{gb2}
    \fmf{dashes,label=$h$,l.dist=2pt}{tth1,tth2}
    \fmf{fermion}{tth2,gb1}
    \fmf{fermion}{gb2,tth2}
    \fmflabel{$b$}{gb1}
    \fmflabel{$\bar{b}$}{gb2}
    \end{fmfgraph*}
    }
    \end{fmffile}
    \label{fig:sig_fd}}
    ~
    \quad\quad\quad
    ~
    \subfloat[Background process.]{
    \begin{fmffile}{ttbb}
    \fmfframe(0,10)(0,17){
    \begin{fmfgraph*}(150,100)
    \fmfleft{spacel1,i2,i1,spacel2}
    \fmfright{spacer1,p5,p3,spacer2}
    \fmftop{space11,space12,space13,p4,p1}
    \fmfbottom{space21,space22,space23,p6,p2}
    \fmf{gluon}{gg1,i1}
    \fmf{gluon}{i2,gg2}
    \fmf{fermion,tension=2}{gg2,gg1}
    \fmf{fermion,label=$t$,l.side=left,tension=2}{gg1,s134}
    \fmf{fermion,label=$\bar{t}$,l.side=left,tension=2}{s256,gg2}
    \fmf{boson,label=$W^+$,l.side=right,tension=1.5}{s134,s13}
    \fmflabel{$b$}{p4}
    \fmf{fermion}{s134,p4}
    \fmf{boson,label=$W^-$,l.side=left,tension=1.5}{s256,s25}
    \fmflabel{$\bar{b}$}{p6}
    \fmf{fermion}{p6,s256}
    \fmflabel{$\ell^+$}{p3}
    \fmf{fermion}{p3,s13}
    \fmflabel{$\ell^-$}{p5}
    \fmf{fermion}{s25,p5}
    \fmflabel{$\nu$}{p1}
    \fmf{fermion}{s13,p1}
    \fmflabel{$\bar{\nu}$}{p2}
    \fmf{fermion}{p2,s25}
    \fmffreeze
    \fmftop{grad1,grad2,gb1,gb2}
    \fmfforce{(0.15w,0.71h)}{grad1}
    \fmfforce{(0.3w,0.8h)}{grad2}
    \fmfforce{(0.33w,0.96h)}{gb1}
    \fmfforce{(0.45w,0.96h)}{gb2}
    \fmf{gluon}{grad2,grad1}
    \fmf{fermion}{grad2,gb1}
    \fmf{fermion}{gb2,grad2}
    \fmflabel{$b$}{gb1}
    \fmflabel{$\bar{b}$}{gb2}
    \end{fmfgraph*}
    }
    \end{fmffile}
    \label{fig:bkg_fd}}
    \caption{Representative LO Feynman diagrams for the signal and background processes used in the $tth$ study.}
    \label{fig:feynman_diagrams}
\end{figure*}
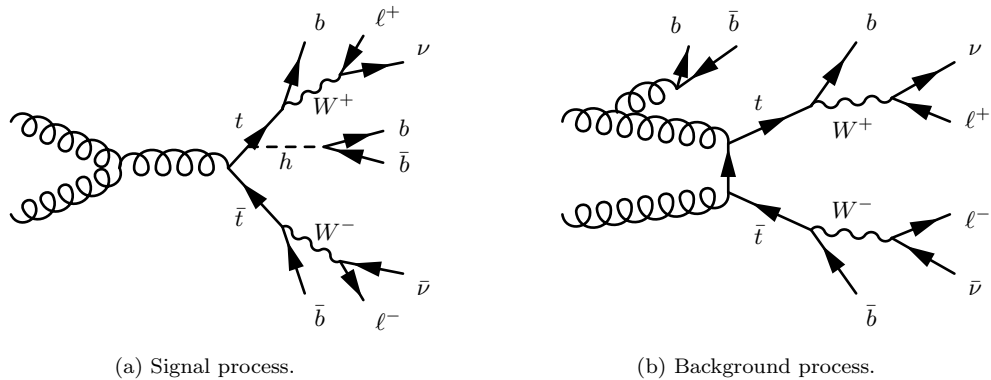

The matrix-element calculations for the signal and background hypotheses were created at leading order (LO) using \texttt{MG5aMC}.
The signal process is $pp\to t\bar{t}h$, where $h\to b\bar{b}$, $t\to W^+b$ and $W^+\to\ell^+\nu$ (and similarly for $\bar{t}$).
Here, $\ell$ refers to the light leptons $e$ and $\mu$.
A representative Feynman diagram for this process is shown in \Cref{fig:sig_fd}.
The background process is $pp\to t\bar{t}b\bar{b}$, where the top quarks decay leptonically as with the signal process.
A representative Feynman diagram for the background is shown in \Cref{fig:bkg_fd}.
Note that in both cases the process can be quark or gluon initiated, and the relative fraction of their density in the proton is calculated with the \texttt{NNPDF23\_lo\_as\_0130\_qed} parton density functions \cite{Ball:2013hta} as implemented in \texttt{LHAPDF}~\cite{Buckley:2014ana}.
In the calculation of the objective function in \Cref{eqn:maximisation}, the matrix-element squared is a weighted sum of each different initial state flavour for the same phase space point, weighted by the corresponding parton density function.
The important parameters in the matrix-element are the masses and widths of the resonances, and were set as follows: $m_t=173$~GeV, $\Gamma_t=1.5$~GeV, $m_W=80.419$~GeV, $\Gamma_W=2.0476$~GeV, $m_h=125$~GeV and $\Gamma_h=6.382$~MeV.

The same matrix-element calculations were also used to generate events for study, with a centre of mass energy of 13~TeV.
In order to account for detector resolution effects, a smearing was applied to the $b$-quarks whereas the lepton four-momenta were assumed to be well measured by the detector.
The smearing modifies the energy of the $b$-quark by a scaling factor and then re-scales the entire three-momentum accordingly, such that the mass remains constant.
This scaling was applied to each $b$-quark in each event, and the scale factor was sampled from a Gaussian distribution with a mean of 1 and width of 0.15.
The precise value of the width was determined by comparing the invariant mass spectrum of the Higgs boson in the signal events with that of the recent ATLAS $h\to b\bar{b}$ result~\cite{Aaboud:2018zhk}.
It was found that a Gaussian resolution of 15\% reproduced the ATLAS simulation best.

Gaussian transfer functions, which mimic the energy smearing described above, were applied to all four of the $b$-quarks.
The transfer function therefore takes the form,
\begin{equation}
    W(x,y)=\prod_{i=1}^4\frac{1}{\sqrt{2\pi }RE_{i}^{x}}\exp\left(-\frac{1}{2}{\left[\frac{E_{i}^{x}-E_{i}^{y}}{RE_{i}^{x}}\right]^2}\right)
    \label{eqn:transfer_function},
\end{equation}
where here $R$ specifies the resolution (15\% in our case), and $E_{i}^x$ and $E_{i}^y$ are the energies of the $i$-th $b$-quark in the event $x$ and in the test point for the maximisation $y$, respectively.
The transfer function therefore controls the maximisation procedure and ensures that the $b$-quark energies remain close to what is recorded in the event $x$.

For both the signal and background hypothesis, there will be 8 unknown quantities in any given event $x$.
These are 3 of the components for each neutrino four-momentum, and the momentum fractions of the 2 incoming partons (the Bjorken-fractions).
Given that all the other final state particles are visible and should contain full information about their four-momentum, one can solve for 4 unknown quantities by requiring an overall conservation of four-momentum for the event, leaving 4 free parameters.
Solutions are only considered usable if they are real, the energies are all positive, and the Bjorken-fractions are between 0 and 1.
In addition to this, the $b$-quarks are assumed to be indistinguishable, and therefore all distinct permutations of the $b$-quarks are tested too, the maximum of which is considered in the maximisation procedure.
The leptons are able to be identified by their flavour and charge, and are therefore not permuted.

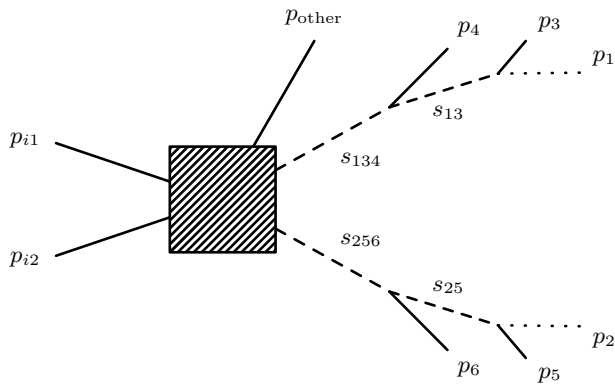
\begin{figure}
    \centering
    \begin{fmffile}{classD}
    \fmfframe(0,10)(0,17){
    \begin{fmfgraph*}(200,120)
    \fmfleft{space01,i2,i1,space02}
    \fmfright{p5,p3}
    \fmftop{space11,space12,pother,p4,p1}
    \fmfbottom{space21,space22,space23,p6,p2}
    \fmflabel{$p_{i1}$}{i1}
    \fmflabel{$p_{i2}$}{i2}
    \fmf{plain}{i1,block}
    \fmf{plain}{i2,block}
    \fmfv{decor.shape=square,decor.filled=shaded,decor.size=0.2w}{block}
    \fmf{plain,tension=0}{block,pother}
    \fmflabel{$p_{\textrm{other}}$}{pother}
    \fmf{dashes,label=$s_{134}$,l.side=right}{block,s134}
    \fmf{dashes,label=$s_{256}$,l.side=left}{block,s256}
    \fmf{dashes,label=$s_{13}$,l.side=right}{s134,s13}
    \fmflabel{$p_4$}{p4}
    \fmf{plain}{s134,p4}
    \fmf{dashes,label=$s_{25}$,l.side=left}{s256,s25}
    \fmflabel{$p_6$}{p6}
    \fmf{plain}{s256,p6}
    \fmflabel{$p_3$}{p3}
    \fmf{plain}{s13,p3}
    \fmflabel{$p_5$}{p5}
    \fmf{plain}{s25,p5}
    \fmflabel{$p_1$}{p1}
    \fmf{dots}{s13,p1}
    \fmflabel{$p_2$}{p2}
    \fmf{dots}{s25,p2}
    \end{fmfgraph*}
    }
    \end{fmffile}
    \caption{The \textit{Main Block D} kinematic template, which can be used to map the unknown invisible particle information, $x_1$ and $x_2$ to the propagator invariant masses both for the signal and background processes.
    }
    \label{fig:classD_kinematics}
\end{figure}

As stated in \Cref{sec:method}, the maximisation procedure is made more efficient by making a transformation to a set of maximisation variables that are peaked at an approximately known location.
For both the signal and background hypothesis, this can be achieved by mapping the four remaining free parameters to the four invariant mass squared variables of the top quark and $W$ boson propagators.
This corresponds with the \textit{Main Block D} computation used in \texttt{MadWeight}~\cite{Artoisenet:2010cn}.
The kinematic template used is illustrated in \Cref{fig:classD_kinematics}.
By solving the requirement of conservation of four-momentum,
\begin{equation}
    p_{i1}^\mu+p_{i2}^\mu=\sum_{i=1}^6p_i^\mu+p_{\text{other}}^\mu,
\end{equation}
and re-casting the final state momenta to their originating propagators,
\begin{align}
    s_{13}&=(p_1^\mu + p_3^\mu)^2, \\
    s_{134}&=(p_1^\mu + p_3^\mu + p_4^\mu)^2, \\
    s_{25}&=(p_2^\mu + p_5^\mu)^2, \\
    s_{256}&=(p_2^\mu + p_5^\mu + p_6^\mu)^2, 
\end{align}
one can recover up to 4 different solutions for the invisible four-momenta and Bjorken-fractions by using the propagator masses as inputs.
During the maximisation procedure, each of these solutions is tested, and the one which returns the highest value for the objective function is chosen.
Comparing \Cref{fig:classD_kinematics} with the final state of interest in this work, we made the following assignments:
\begin{align*}
    &p_1\leftrightarrow\nu_1,\quad p_2\leftrightarrow\nu_2, \\
    &p_3\leftrightarrow\ell_1,\quad p_5\leftrightarrow\ell_2, \\
    &p_4\leftrightarrow b_1,\quad p_6\leftrightarrow b_2, \\
    &p_{\text{other}}\leftrightarrow b_3+b_4.
\end{align*}
For the signal hypothesis, $p_{\text{other}}$ corresponds to the four-momentum of the Higgs boson, whereas for the background it merely corresponds to the additional $b$-quarks not produced by the top decays.

For both the signal and the background, the four common propagator masses were used as variables in the maximisation.
That is,
\begin{align*}
    &\sqrt{s_{134}}\leftrightarrow m_{t1},\quad \sqrt{s_{256}}\leftrightarrow m_{t2}, \\
    &\sqrt{s_{13}}\leftrightarrow m_{W1},\quad \sqrt{s_{25}}\leftrightarrow m_{W2}.
\end{align*}
For the background hypothesis, another four variables are added to the maximisation procedure, namely the energy scaling of each of the $b$-quarks.
This is necessary to allow the transfer function to be constrained.
The signal hypothesis is slightly different, and has the additional constraint that the invariant mass of the $b_3$-$b_4$ system is equal to the Higgs mass $m_h$.
Therefore, the four-momentum of $b_3$ is fixed as a result of the constraint, and instead the Higgs mass is used as a variable in the maximisation.
This greatly improves the efficiency when considering a signal-like event, since the Higgs boson has a narrow width.

Finally, the maximisation variables are further mapped to a region where they are distributed in the region between 0 and 1.
This is accomplished by constructing the cumulative density function of the Breit-Wigner probability density function (PDF) for all of the mass related variables, and that of the Gaussian PDF for the energy scaling variables.
The parameters used for the Breit-Wigners are the known masses and widths of the resonances, and for the Gaussians the central energy and resolution are used as the mean and width (as in the transfer function in \Cref{eqn:transfer_function}).
This mapping has no physical consequences, it is merely used to enhance the efficiency of the maximisation procedure by allowing it a more evenly distributed parameter space to optimise.

The maximisation procedure also has a number of parameters that can be set, which determine the quality of the output.
The most important of these are the stopping criteria for the algorithms, which include the threshold values of the objective function and maximisation parameters as well as a cut-off time.
In the control set-up, the algorithm was set to stop if it could find a maximum to within 1\% precision in the objective function or 0.1\% precision in the maximisation variables.
If the algorithm fails to find a maximum with this precision, it will terminate after 200 seconds of operation. 

\subsection{Performance tests}
\label{sec:tests}

The control set-up described in \Cref{sec:control} was used as a baseline to undergo a few performance tests with regards to the different optimisation algorithms and parameters of the method.
First, a test of the CPU efficiency for each derivative-free optimisation algorithm shown in \Cref{tab:nlopt_algorithms} was made.
Thereafter, the performance of the algorithms was compared with the metric of discovery significance per inverse femtobarn of $\sqrt{s}=13$~TeV data at the LHC.

A scan of the different optimisation algorithms was performed using a total of 2000 input events, half of them signal events and the other half background events.
The CPU time per event was recorded for maximising the event under both the signal and background hypotheses.
The results of this study are shown in \Cref{tab:algorithm_times}.
Also shown here is the percentage of results which returned a ``zero''.
In general, if the algorithm returned a zero as the maximum of the objective function, it is due to a failure of the algorithm.
Algorithms in \Cref{tab:nlopt_algorithms} that do not feature in \Cref{tab:algorithm_times} are those that on average took more than the maximum time of 200 seconds per event, and were therefore discarded from further tests.

\begin{table*}
    \centering
    \setlength{\tabcolsep}{8pt}
    \begin{tabular}{l|ccc|cc}
    \multirow{2}{*}{\textbf{Algorithm}} & \multicolumn{3}{c}{\textbf{Time per event $\pm$ standard deviation [s]}} & \multicolumn{2}{|c}{\textbf{Zero results [\%]}} \\
    \cline{2-6}
    &            \textbf{Background} &        \textbf{Signal} &       \textbf{Total} &        \textbf{Background} &        \textbf{Signal} \\
    \hline
    \texttt{GN\_CRS2\_LM} &           43.95~$\pm$~27.92 &      107.17~$\pm$~42.18 &     151.12~$\pm$~53.57 &            0.10 &        0.10 \\
    \texttt{GN\_DIRECT} &           14.18~$\pm$~23.11 &       24.53~$\pm$~41.25 &      38.71~$\pm$~51.96 &            0.95 &        3.05 \\
    \texttt{GN\_DIRECT\_L} &           15.63~$\pm$~33.07 &       22.86~$\pm$~39.15 &      38.49~$\pm$~56.08 &            1.70 &        4.20 \\
    \texttt{GN\_DIRECT\_L\_NOSCAL} &           13.10~$\pm$~26.33 &       20.85~$\pm$~40.54 &      33.95~$\pm$~54.52 &            1.65 &        4.55 \\
    \texttt{GN\_DIRECT\_L\_RAND} &            5.33~$\pm$~14.93 &       12.66~$\pm$~36.91 &      18.00~$\pm$~42.43 &            0.35 &        3.30 \\
    \texttt{GN\_DIRECT\_L\_RAND\_NOSCAL} &           10.86~$\pm$~28.71 &       17.78~$\pm$~39.43 &      28.64~$\pm$~53.30 &            1.60 &        4.25 \\
    \texttt{GN\_DIRECT\_NOSCAL} &           10.56~$\pm$~25.75 &       17.92~$\pm$~41.40 &      28.48~$\pm$~54.76 &            1.60 &        4.55 \\
    \texttt{GN\_ORIG\_DIRECT\_L} &           48.47~$\pm$~25.29 &       51.42~$\pm$~33.94 &      99.89~$\pm$~46.93 &            0.45 &        3.75 \\
    \texttt{LN\_AUGLAG} &            3.18~$\pm$~2.13 &        4.63~$\pm$~4.86 &       7.81~$\pm$~5.50 &            9.00 &       12.40 \\
    \texttt{LN\_AUGLAG\_EQ} &            3.12~$\pm$~2.02 &        4.57~$\pm$~5.62 &       7.69~$\pm$~6.12 &            9.05 &       12.85 \\
    \texttt{LN\_BOBYQA} &            1.69~$\pm$~0.85 &        2.61~$\pm$~1.32 &       4.30~$\pm$~1.78 &            6.10 &        8.10 \\
    \texttt{LN\_COBYLA} &            5.15~$\pm$~3.77 &        7.33~$\pm$~8.83 &      12.48~$\pm$~9.87 &            9.35 &       12.60 \\
    \texttt{LN\_NELDERMEAD} &            6.85~$\pm$~3.68 &       11.39~$\pm$~6.35 &      18.23~$\pm$~8.16 &            8.35 &       10.35 \\
    \texttt{LN\_NEWUOA} &            1.68~$\pm$~0.76 &        3.16~$\pm$~1.80 &       4.84~$\pm$~2.05 &            6.05 &        8.40 \\
    \texttt{LN\_NEWUOA\_BOUND} &            1.68~$\pm$~0.84 &        1.92~$\pm$~0.69 &       3.60~$\pm$~1.40 &            6.15 &        8.45 \\
    \texttt{LN\_PRAXIS} &            5.87~$\pm$~3.20 &       10.79~$\pm$~8.67 &      16.67~$\pm$~9.26 &            5.75 &        8.30 \\
    \texttt{LN\_SBPLX} &            8.86~$\pm$~5.57 &       14.73~$\pm$~11.42 &      23.59~$\pm$~13.69 &            6.70 &        9.20 \\
    \hline
\end{tabular}
    \caption{Results for the algorithm scan, which tested each algorithm in terms of its time per event for both the signal and background hypotheses.
    Also shown is the percentage of results which found a maximum of zero,  which usually corresponds to a failure of the algorithm to find a maximum.}
    \label{tab:algorithm_times}
\end{table*}

From the results in \Cref{tab:algorithm_times}, it is clear that the local algorithms (i.e. those that start with ``\texttt{LN}'') take significantly quicker on average to find a maximum than the global algorithms.
This is expected, and not necessarily a clear reason to favour the local over the global algorithms, since it is far more likely that the local algorithms find a local maximum as opposed to a global maximum.
The local algorithms are also far more likely to fail and return a zero result.
In all cases, the signal hypothesis appears to take longer than the background hypothesis.
This decrease in speed is solely due to the fact that maximising the background events with the signal hypothesis is the most computationally difficult problem for the maximisation algorithm to solve.
The algorithm has the constraint that the invariant mass of the $b_1$-$b_3$ system needs to be close to the mass of the Higgs boson, and for background events it can be difficult to find a phase space point which satisfies this without varying the transfer function substantially.

Now, in order to determine which algorithm performs best, a comparative study of the approximate expected discovery significance was conducted.
This was done as follows.
For each event, the discriminant $\chi$ in \Cref{eqn:chi_discriminant} is calculated with information about the event source.
Events which are more signal-like have higher values in $\chi$, so a lower cut $\chi_0$ is made to the distribution, thereby introducing the acceptance factors $\mathcal{A}_s$ and $\mathcal{A}_b$ for the signal and background events, respectively.
Note that events with a zero background weight are subtracted from the background acceptance factor, and events where both the signal and background weight are zero are ignored.
The optimal cut is that which maximises the Poisson significance,
\begin{equation}
    Z=\frac{N_s}{\sqrt{N_s+N_b}},
    \label{eqn:significance}
\end{equation}
where the number of events can be determined as follows,
\begin{equation}
    N_s=L\sigma_s\mathcal{A}_s\epsilon_s,
    \label{eqn:num_events}
\end{equation}
and similarly for $N_b$.
Here the luminosity $L$ is set to 1~fb$^{-1}$, but it is clear to see that the significance will scale as the square root of $L$.
The signal cross section $\sigma_s$ was taken from the LHC Higgs cross section working group~\cite{deFlorian:2016spz} and the background cross section $\sigma_b$ was calculated at next-to-leading order in \texttt{MG5aMC}.
After being multiplied by the appropriate branching ratios, these numbers were found to be $\sigma_s=13.45$~fb and $\sigma_b=723.62$~fb.
The efficiencies $\epsilon_s$ and $\epsilon_b$ relate to the experimental efficiency of reconstructing the final state, and were both set to an approximated value of 2.5\%.\footnote{Note that the goal of this study is not to accurately calculate the significance; these numbers merely provide an estimate that can be used to make comparisons between the different algorithms.}

\begin{figure}
    \centering
    \includegraphics[width=0.5\textwidth]{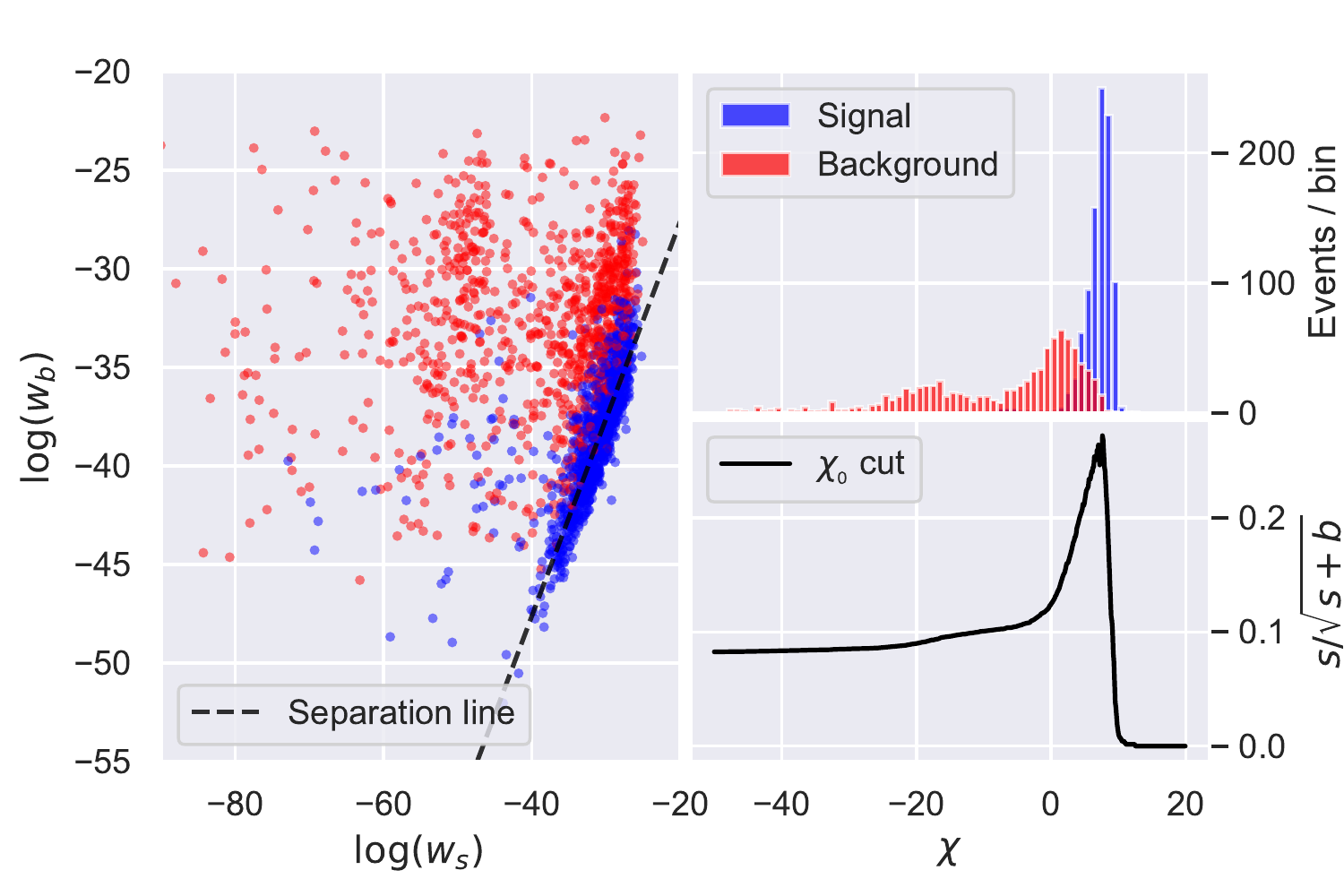}
    \caption{An example of the significance calculation for the \texttt{GN\_DIRECT\_L\_RAND} algorithm.
    On the left, a scatter plot of the log of the weights from the maximisation procedure.
    On the upper-right, a distribution of the discriminant $\chi$, and below an illustration of how the optimal cut value is found.
    Signal events are shown in blue and background events in red.}
    \label{fig:GN_DIRECT_L_RAND}
\end{figure}

An example of this calculation for the \texttt{GN\_DIRECT\_L\_RAND} algorithm is shown in \Cref{fig:GN_DIRECT_L_RAND}.
On the left is shown a scatter plot of the signal hypothesis weights on the horizontal axis compared with the background hypothesis weights on the vertical axis.
The events are colour-coded by their source, and so the separation between the signal and background events is apparent.
The distribution for the discriminant $\chi$ is shown in the upper right again with information about the event source.
Below this plot is the value of the significance with $L=1$~fb$^{-1}$ as a function of the cut value $\chi_0$.
The optimal cut value is then used to draw the separation line on the left.
As mentioned in \Cref{sec:method}, this one-dimensional separation method is not optimal but is good enough as a baseline for the study.
It is also interesting to note that the signal events are directly correlated when looking at the scatter plot, thereby indicating that the weights do indeed behave like (un-normalised) probabilities.

\begin{figure}
    \centering
    \includegraphics[width=0.5\textwidth]{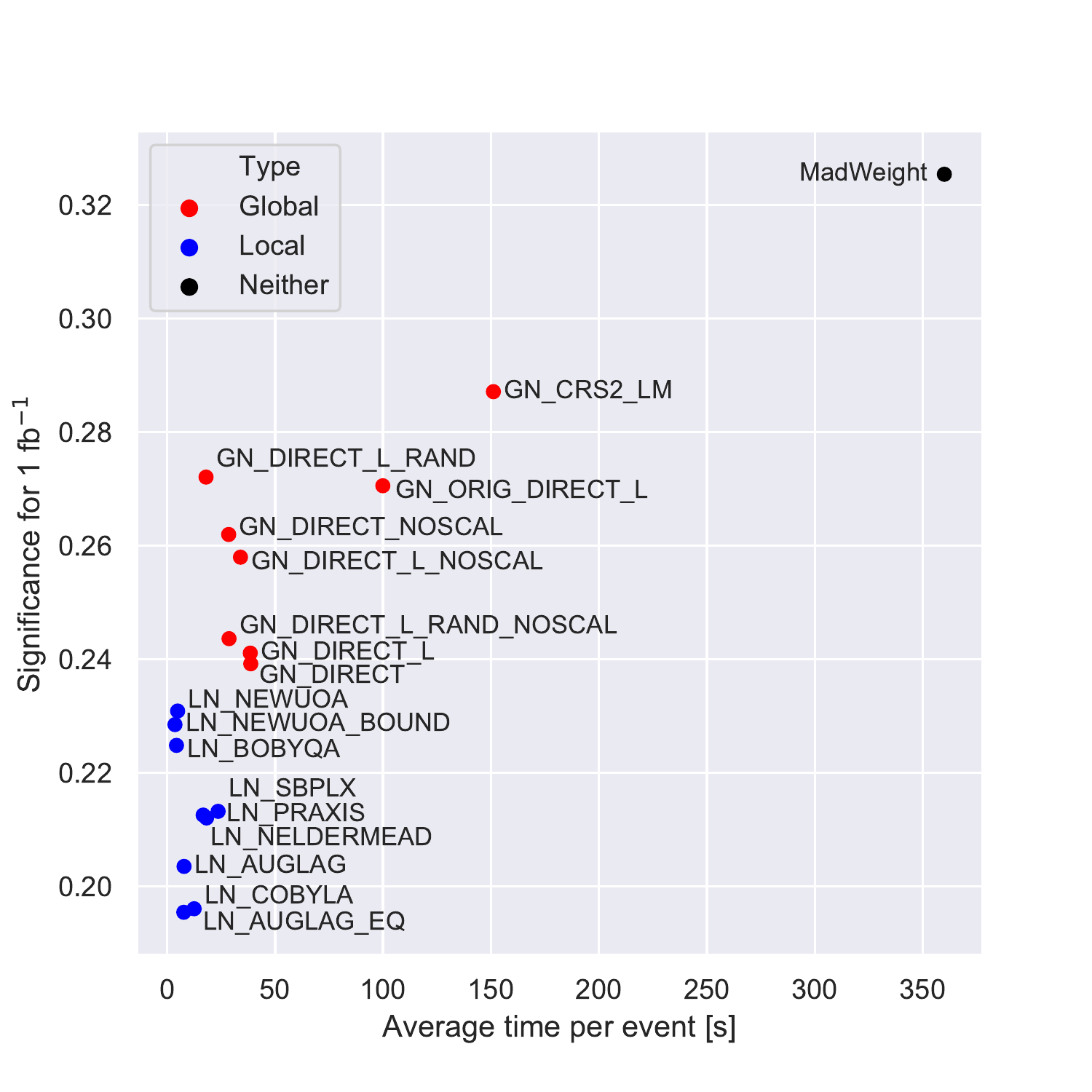}
    \caption{The results from the comparative study of the performance for the different maximisation algorithms, shown in terms of average time per event and the expected discovery significance.
    The maximisation algorithms are compared with the traditional MEM calculation, as shown by the black dot.}
    \label{fig:significance_time}
\end{figure}

The results of the comparative study for all the algorithms that succeeded is shown in \Cref{fig:significance_time}.
On the horizontal axis is the average time per event, and on the vertical axis is the significance produced for each algorithm, calculated as described above.
It is reassuring to note that the global algorithms all perform better that the local algorithms, since they are more likely to find a global maximum.
It is also apparent that it should cost more CPU time to make a better quality discriminant, although in a few cases it is not obvious.
The maximisation algorithms are also compared to the traditional MEM as implemented by \texttt{MadWeight}.
As expected, the traditional MEM does a better job overall in performing a separation between the signal and background events at the cost of a significantly higher time per event.
It should also be noted that \texttt{MadWeight} had been under development for several years, and so was optimised to run as efficiently as possible.
The maximisation procedure described in this work has yet to be optimised.
Therefore, it could be argued that the maximisation procedure could be significantly more CPU-efficient than what is shown here. 
The traditional MEM as implemented in \texttt{MadWeight} has a $15\%$ higher significance for the example at hand, compared to the best optimiser applying the maximisation method, at the cost of $2.5$ times more time needed to evaluate an event.

Choosing an ``optimal'' algorithm depends on how much one values the separation power compared with the average time per event.
If speed is of utmost importance, the optimal algorithm is \texttt{LN\_NEWUOA\_BOUND}, whereas the \texttt{GN\_CRS2\_LM} algorithm has the best separation power.
For a healthy balance of both factors, it would appear that the \texttt{GN\_DIRECT\_L\_RAND} algorithm seems to be the optimal choice.
Therefore, for the rest of this paper, the algorithm of choice shall be \texttt{GN\_DIRECT\_L\_RAND}.

\subsection{Further discussion}

In view of the success of the maximisation procedure explored in this work, one might ask the whether or not it has any utility beyond being used for a classifier in particle physics searches.
In particular, the maximisation procedure can provide solutions for the four-momenta of the invisible particles defined for a given process.
If one could estimate the four-momenta of the invisible particles in an input event, it would help accelerate the difficulty of searching for final states with complicated decay chains involving invisible particles.
In order to understand whether matrix-element maximisation can provide a suitable approximation for the four-momenta of multiple invisible particles, a few tests were performed.

As in \Cref{sec:tests}, a set of 1000 signal events (with the 15\% energy smearing of the $b$-quarks) was used to perform the maximisation procedure with the \texttt{GN\_DIRECT\_L\_RAND} algorithm.
However, in this case, the input event was saved with full information about the momenta of the two neutrinos, and was then compared with the event that resulted from the maximisation procedure.
Using the signal events and the signal hypothesis, it was then possible to compare the ``actual'' values of $p_\text{T}$ and $p_Z$ of the neutrinos with the values that maximise the objective function in \Cref{eqn:maximisation}. 
As a result of this, it was found that there is a non-trivial bias between the two different cases.
On average, the reconstructed $p_\text{T}$ values of the neutrinos from the maximisation were about 70\% larger than the ``actual'' values in the input event with a standard deviation of up to 350\%.
The difference in the values of $p_Z$ were smaller on a direct comparison, changing only by up to about 3~GeV on average.
However, the standard deviation on this difference could be as large as 235~GeV.

\begin{figure*}
    \centering
    \subfloat[Signal events with the signal hypothesis.]{
        \includegraphics[width=1.045\textwidth]{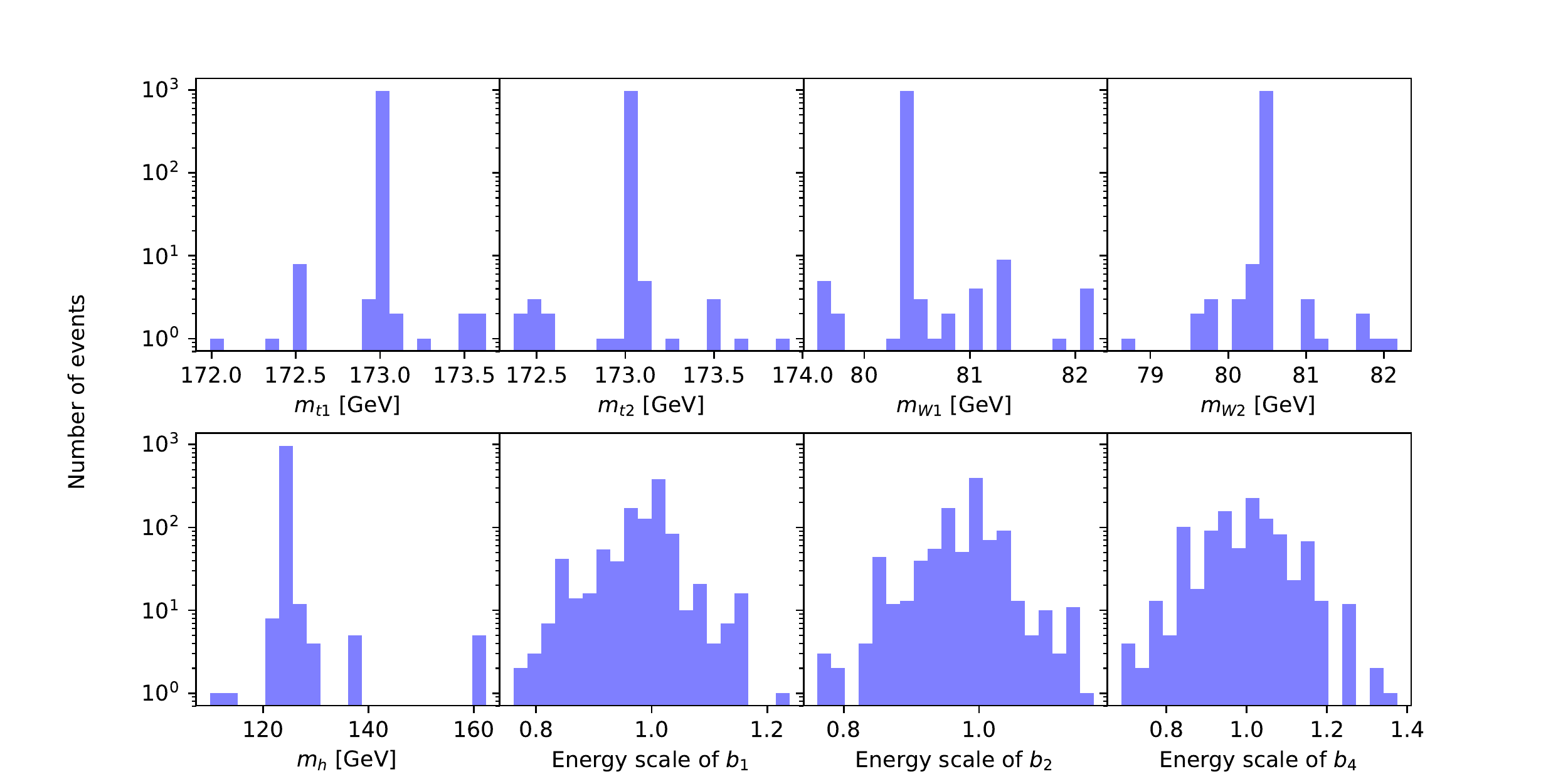}
    }
    
    \subfloat[Background events with the background hypothesis.]{
    \includegraphics[width=1.045\textwidth]{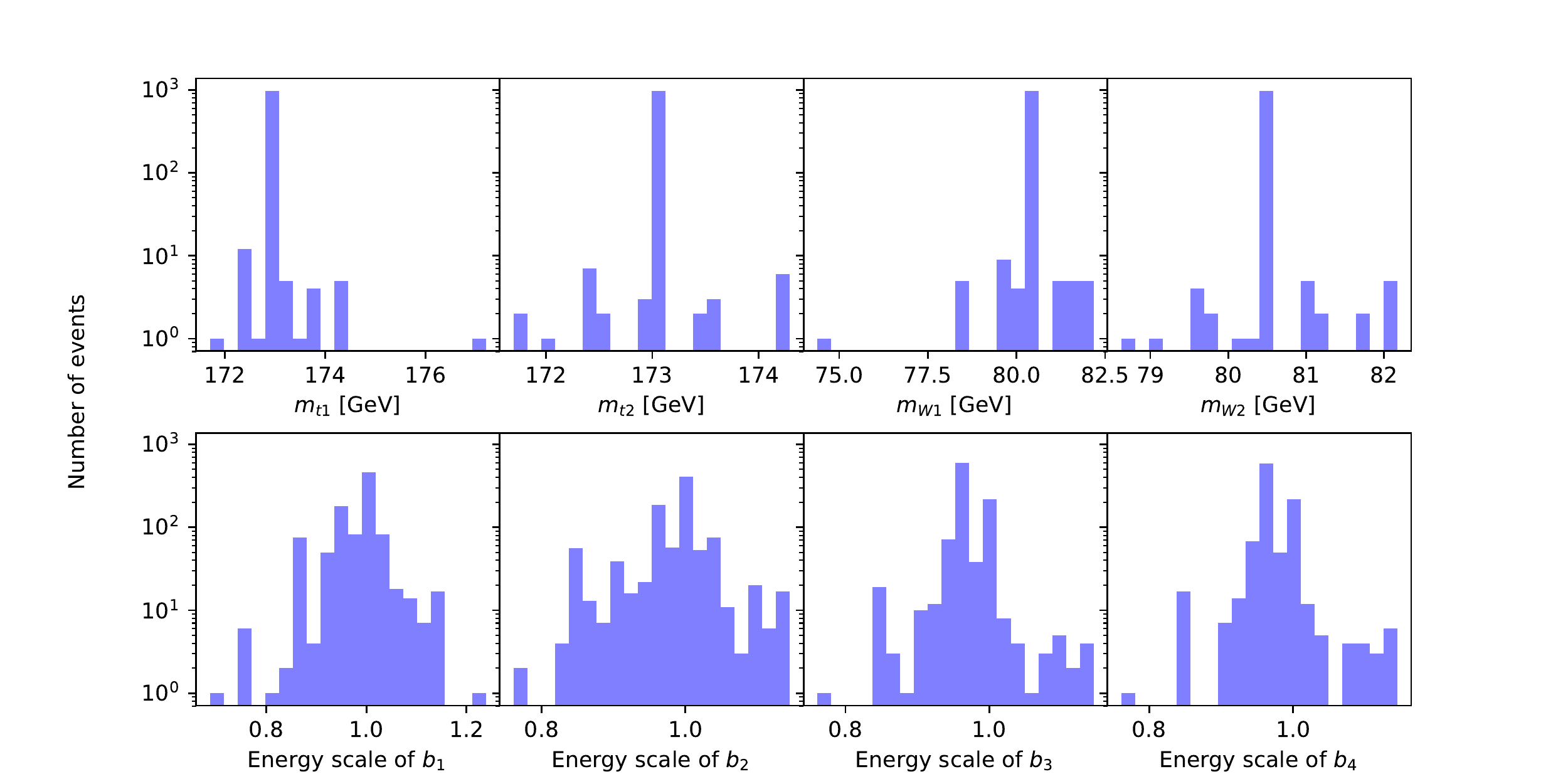}
    }
    \caption{Distributions of the maximisation variables at the values which maximise the objective function in \Cref{eqn:maximisation}, using the \texttt{GN\_DIRECT\_L\_RAND} algorithm.
    The choice of these variables is described in \Cref{sec:control}.}
    \label{fig:reconstructed_kinematics}
\end{figure*}

The maximised events therefore do not provide a precise estimate of the neutrino four-momenta, but come close on average.
The driving reason for this is that the matrix-element squared is driven by the momentum of the propagators.
For the maximisation algorithm, it is far more efficient to find a maximum by setting the mass-like parameters to the pole masses of the intermediate particles, and then allowing the transfer function to vary such that the $b$-quarks can satisfy the appropriate constraints to maximise the objective function.
This can be seen in \Cref{fig:reconstructed_kinematics}.
Whereas any production of such a process would result in a Breit-Wigner distribution for the propagator masses, the distribution of the maximised masses is far from a Breit-Wigner, where here the pole masses are preferred.
This does indeed result in the maximisation of the objective function, however the four-momenta of the neutrinos are then forced away from their ``actual'' values.

\begin{figure}
    \centering
    \includegraphics[width=0.5\textwidth]{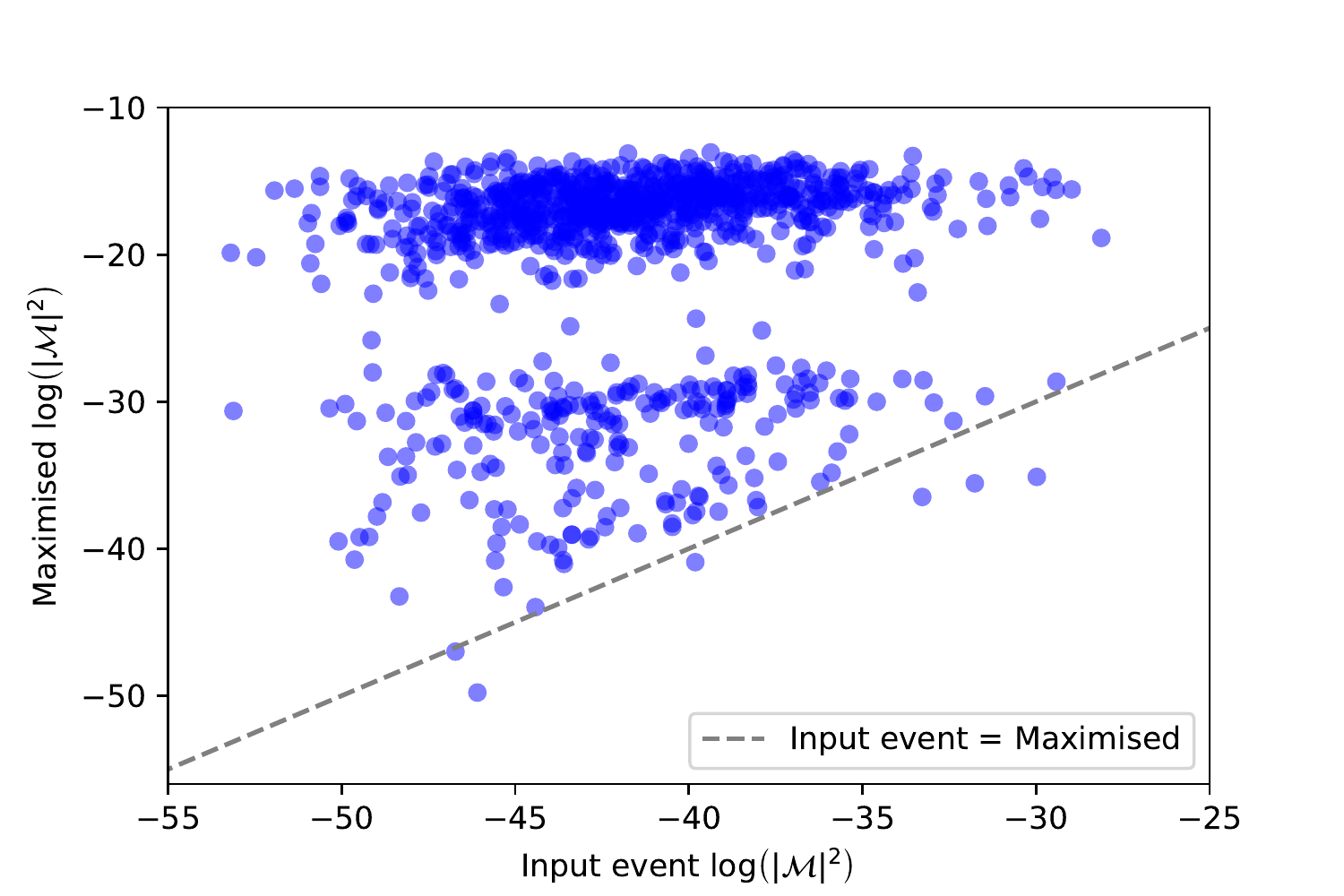}
    \caption{The comparison of the ``actual'' value of the matrix-element squared with that of the maximised events (shown here without multiplication of the transfer function).
    This plot includes only signal events, and the maximisation algorithm used is the \texttt{GN\_DIRECT\_L\_RAND} algorithm.
    The dotted line shows where the two cases are equivalent.}
    \label{fig:max_lhe_ME}
\end{figure}

It can also be shown that the maximised value of the matrix-element squared is not correlated with that of the input event in any case.
In \Cref{fig:max_lhe_ME} is shown a scatter plot of the matrix-element squared values for the input event (with all information about the neutrinos) compared with that of the events after the maximisation procedure.
Note that these point do not include multiplication by the transfer function.
It is apparent that the value of $\log\left|\mathcal{M}\right|^2$ for the maximised events is not at all sensitive to the same value calculated with the input events.
Another interesting point is that in very few cases, the maximised value ends up smaller than the ``actual'' value.
This points to a deficiency in the maximisation algorithm, which is expected when looking at \Cref{fig:significance_time}.

In general, the maximisation method will always provide a well defined set of four-momenta for any invisible particles defined in the matrix-element.
From the studies shown above, it can be argued that these four-momenta will be a more accurate representation of the ``real'' invisible momenta when the resonances in question have a narrow width.
This is because the matrix-element is dominated by the shape of the Breit-Wigner distributions for each resonance, and the maximum is almost always found at their Breit-Wigner peaks.
Therefore, narrow width resonances (which will always be produced with a mass close to their pole masses) will return a set of invisible four-momenta that is representative of the ``real'' invisible momenta.
This is an important fact to keep in mind for future studies, where the matrix-element maximisation method can be used as a CPU-efficient guess of the invisible four-momenta, and these momenta can be further used as a seed for more complicated MEMs (such as Event Deconstruction~\cite{Soper:2014rya}) when dealing with multiple invisible particles.
The intrinsic shortcoming of the maximisation method is that the guess will not always be accurate when large width resonances are considered.

\section{Summary \& conclusions}
\label{sec:summary}

In this article, we have presented a detailed study of the matrix-element maximisation method of signal processing as applied to the $tth$ process.
The primary motivation for the development of this technique is to serve as a MEM that is more CPU-efficient than the traditional integration-based MEM.
An interesting consequence of using the maximisation method is that one can make an educated guess for the four-momenta of an arbitrary number of invisible particles defined in the process in question.
This guess can be used in different signal processing techniques that require these momenta as an input.
This work therefore serves two purposes, one being to prove that the maximisation technique is a decent classifier and improves on the CPU-efficiency of the traditional MEM.
The other purpose is as an intermediate work towards a generic method of obtaining invisible momenta for more complicated MEMs, e.g. Shower or Event Deconstruction.

The primary motivation of this work has been shown through a comprehensive study of the maximisation method using a variety of different maximisation algorithms.
In Ref.~\cite{FerreiradeLima:2017iwx}, it had already been shown that the matrix-element maximisation technique does indeed act as a suitable classifier for particle physics searches.
In this work, however, this has been quantified in terms of how well the method performs (in terms of discovery significance) as well as how CPU-efficient it is (in terms of CPU time per event).
Depending on the algorithm used, it has been shown that the maximisation method can improve on the speed of the traditional MEM by up to two orders of magnitude.
Even so, the procedure used in this work has not been optimised, and so speed improvements can be expected in future developments.
The cost of this speed improvement is a decrease in the quality of the classifier, which performs between about 60\%-85\% of the traditional MEM\footnote{The traditional, or full, matrix-element method integrates the matrix-element over the entire available phase space of invisible particles, thereby summing up the matrix-element weights over all kinematically allowed phase space regions, instead of selecting only the maximum matrix-element weight possible for a given final state.} in terms of discovery significance.
Should the method be used in practice, the user has a range of algorithm choices to consider depending on how much they value speed compared with performance.
In this work, it was found that the \texttt{GN\_DIRECT\_L\_RAND} algorithm is a good choice for a balanced consideration of both speed and performance.

The consequence of the maximisation method's ability to provide a guess of the invisible particles' four-momenta has also been discussed.
As a conclusion, it has been found that the maximisation procedure will most often find a maximum at the pole masses of the resonances in question.
At this point, the four-momenta of the invisible particles is defined.
Therefore, the four-momenta will more often be an accurate representation of the ``true'' value when the resonances have a narrow width.
For larger width resonances, one could attempt to quantify the expected uncertainty of the invisible momenta and use it as a systematic uncertainty; this is left as a future study.
The resulting guess of the invisible momenta can be used as a starting point for more complicated techniques.

At the endpoint of this work, the underlying goal of the method has shifted towards a new focal point.
As a future study, the maximisation method discussed here can be implemented into an automated system, combining the ability to guess invisible four-momenta with techniques that attempt to reconstruct entire event histories.
This new goal is evidently obtainable without too many obstacles, since the automation of the traditional MEM has already been shown to be possible in the past (i.e.~in \texttt{MadWeight}~\cite{Artoisenet:2013vfa}).
It is hoped that the studies done in this article and the proposed developments can contribute to the important goal of finding elusive signals of new physics at the LHC and beyond.

\begin{acknowledgments}
This work has received funding from the European Union's Horizon 2020 research and innovation programme as part of the Marie Sk\l{}odowska-Curie Innovative Training Network MCnetITN3 (grant agreement no. 722104).
MS would like to thank the Humboldt Society for support and the University of Tuebingen for hospitality during the finalisation of parts of this work. 
Computational resources have been provided by the Consortium des \'Equipements de Calcul Intensif (C\'ECI), funded by the Fonds de la Recherche Scientifique de Belgique (F.R.S.-FNRS) under Grant No. 2.5020.11 and by the Walloon Region.
\end{acknowledgments}

\appendix


\bibliographystyle{utphys}
\bibliography{references}

\providecommand{\href}[2]{#2}\begingroup\raggedright\begin{thebibliography}{10}

\bibitem{Kondo:1988yd}
K.~Kondo, ``{Dynamical Likelihood Method for Reconstruction of Events With
  Missing Momentum. 1: Method and Toy Models},''
\href{http://dx.doi.org/10.1143/JPSJ.57.4126}{{\em J. Phys. Soc. Jap.}
  {\bfseries 57} (1988) 4126--4140}.

\bibitem{Abazov:2004cs}
{\bfseries D0} Collaboration, V.~M. Abazov {\em et~al.}, ``{A precision
  measurement of the mass of the top quark},''
  \href{http://dx.doi.org/10.1038/nature02589}{{\em Nature} {\bfseries 429}
  (2004) 638--642},
\href{http://arxiv.org/abs/hep-ex/0406031}{{\ttfamily arXiv:hep-ex/0406031
  [hep-ex]}}.

\bibitem{Artoisenet:2010cn}
P.~Artoisenet, V.~Lemaitre, F.~Maltoni, and O.~Mattelaer, ``{Automation of the
  matrix element reweighting method},''
  \href{http://dx.doi.org/10.1007/JHEP12(2010)068}{{\em JHEP} {\bfseries 12}
  (2010) 068},
\href{http://arxiv.org/abs/1007.3300}{{\ttfamily arXiv:1007.3300 [hep-ph]}}.

\bibitem{Albertsson:2018maf}
K.~Albertsson {\em et~al.}, ``{Machine Learning in High Energy Physics
  Community White Paper},''
  \href{http://dx.doi.org/10.1088/1742-6596/1085/2/022008}{{\em J. Phys. Conf.
  Ser.} {\bfseries 1085} no.~2, (2018) 022008},
\href{http://arxiv.org/abs/1807.02876}{{\ttfamily arXiv:1807.02876
  [physics.comp-ph]}}.

\bibitem{Brehmer:2019xox}
J.~Brehmer, F.~Kling, I.~Espejo, and K.~Cranmer, ``{MadMiner: Machine
  learning-based inference for particle physics},''
\href{http://arxiv.org/abs/1907.10621}{{\ttfamily arXiv:1907.10621 [hep-ph]}}.

\bibitem{James:2000et}
F.~James, Y.~Perrin, and L.~Lyons, eds., {\em {Workshop on confidence limits,
  CERN, Geneva, Switzerland, 17-18 Jan 2000: Proceedings}}.
\newblock 2000.
\newblock
\url{http://weblib.cern.ch/abstract?CERN-2000-005}.
\newblock

\bibitem{Cranmer:2006zs}
K.~Cranmer and T.~Plehn, ``{Maximum significance at the LHC and Higgs decays to
  muons},'' \href{http://dx.doi.org/10.1140/epjc/s10052-007-0309-4}{{\em Eur.
  Phys. J.} {\bfseries C51} (2007) 415--420},
\href{http://arxiv.org/abs/hep-ph/0605268}{{\ttfamily arXiv:hep-ph/0605268
  [hep-ph]}}.

\bibitem{Alwall:2010cq}
J.~Alwall, A.~Freitas, and O.~Mattelaer, ``{The Matrix Element Method and QCD
  Radiation},'' \href{http://dx.doi.org/10.1103/PhysRevD.83.074010}{{\em Phys.
  Rev.} {\bfseries D83} (2011) 074010},
\href{http://arxiv.org/abs/1010.2263}{{\ttfamily arXiv:1010.2263 [hep-ph]}}.

\bibitem{Alwall:2009sv}
J.~Alwall, A.~Freitas, and O.~Mattelaer, ``{Measuring Sparticles with the
  Matrix Element},'' \href{http://dx.doi.org/10.1063/1.3327611}{{\em AIP Conf.
  Proc.} {\bfseries 1200} (2010) 442--445},
\href{http://arxiv.org/abs/0910.2522}{{\ttfamily arXiv:0910.2522 [hep-ph]}}.

\bibitem{Andersen:2012kn}
J.~R. Andersen, C.~Englert, and M.~Spannowsky, ``{Extracting precise Higgs
  couplings by using the matrix element method},''
  \href{http://dx.doi.org/10.1103/PhysRevD.87.015019}{{\em Phys. Rev.}
  {\bfseries D87} no.~1, (2013) 015019},
\href{http://arxiv.org/abs/1211.3011}{{\ttfamily arXiv:1211.3011 [hep-ph]}}.

\bibitem{Artoisenet:2013vfa}
P.~Artoisenet, P.~de~Aquino, F.~Maltoni, and O.~Mattelaer, ``{Unravelling
  $t\overline{t}h$ via the Matrix Element Method},''
  \href{http://dx.doi.org/10.1103/PhysRevLett.111.091802}{{\em Phys. Rev.
  Lett.} {\bfseries 111} no.~9, (2013) 091802},
\href{http://arxiv.org/abs/1304.6414}{{\ttfamily arXiv:1304.6414 [hep-ph]}}.

\bibitem{Betancur:2017kqe}
A.~Betancur, D.~Debnath, J.~S. Gainer, K.~T. Matchev, and P.~Shyamsundar,
  ``{Measuring the mass, width, and couplings of semi-invisible resonances with
  the Matrix Element Method},''
\href{http://arxiv.org/abs/1708.07641}{{\ttfamily arXiv:1708.07641 [hep-ph]}}.

\bibitem{Soper:2011cr}
D.~E. Soper and M.~Spannowsky, ``{Finding physics signals with shower
  deconstruction},'' \href{http://dx.doi.org/10.1103/PhysRevD.84.074002}{{\em
  Phys. Rev.} {\bfseries D84} (2011) 074002},
\href{http://arxiv.org/abs/1102.3480}{{\ttfamily arXiv:1102.3480 [hep-ph]}}.

\bibitem{Soper:2012pb}
D.~E. Soper and M.~Spannowsky, ``{Finding top quarks with shower
  deconstruction},'' \href{http://dx.doi.org/10.1103/PhysRevD.87.054012}{{\em
  Phys. Rev.} {\bfseries D87} (2013) 054012},
\href{http://arxiv.org/abs/1211.3140}{{\ttfamily arXiv:1211.3140 [hep-ph]}}.

\bibitem{deLima:2014dta}
D.~E. Ferreira~de Lima, A.~Papaefstathiou, and M.~Spannowsky, ``{Standard model
  Higgs boson pair production in the ( $ b\overline{b} $ )( $ b\overline{b} $ )
  final state},'' \href{http://dx.doi.org/10.1007/JHEP08(2014)030}{{\em JHEP}
  {\bfseries 08} (2014) 030},
\href{http://arxiv.org/abs/1404.7139}{{\ttfamily arXiv:1404.7139 [hep-ph]}}.

\bibitem{FerreiradeLima:2016gcz}
D.~Ferreira~de Lima, P.~Petrov, D.~Soper, and M.~Spannowsky, ``{Quark-Gluon
  tagging with Shower Deconstruction: Unearthing dark matter and Higgs
  couplings},'' \href{http://dx.doi.org/10.1103/PhysRevD.95.034001}{{\em Phys.
  Rev.} {\bfseries D95} no.~3, (2017) 034001},
\href{http://arxiv.org/abs/1607.06031}{{\ttfamily arXiv:1607.06031 [hep-ph]}}.

\bibitem{Campbell:2012cz}
J.~M. Campbell, W.~T. Giele, and C.~Williams, ``{The Matrix Element Method at
  Next-to-Leading Order},''
  \href{http://dx.doi.org/10.1007/JHEP11(2012)043}{{\em JHEP} {\bfseries 11}
  (2012) 043},
\href{http://arxiv.org/abs/1204.4424}{{\ttfamily arXiv:1204.4424 [hep-ph]}}.

\bibitem{Campbell:2013hz}
J.~M. Campbell, R.~K. Ellis, W.~T. Giele, and C.~Williams, ``{Finding the Higgs
  boson in decays to $Z \gamma$ using the matrix element method at
  Next-to-Leading Order},''
  \href{http://dx.doi.org/10.1103/PhysRevD.87.073005}{{\em Phys. Rev.}
  {\bfseries D87} no.~7, (2013) 073005},
\href{http://arxiv.org/abs/1301.7086}{{\ttfamily arXiv:1301.7086 [hep-ph]}}.

\bibitem{Martini:2015fsa}
T.~Martini and P.~Uwer, ``{Extending the Matrix Element Method beyond the Born
  approximation: Calculating event weights at next-to-leading order
  accuracy},'' \href{http://dx.doi.org/10.1007/JHEP09(2015)083}{{\em JHEP}
  {\bfseries 09} (2015) 083},
\href{http://arxiv.org/abs/1506.08798}{{\ttfamily arXiv:1506.08798 [hep-ph]}}.

\bibitem{Gritsan:2016hjl}
A.~V. Gritsan, R.~Röntsch, M.~Schulze, and M.~Xiao, ``{Constraining anomalous
  Higgs boson couplings to the heavy flavor fermions using matrix element
  techniques},'' \href{http://dx.doi.org/10.1103/PhysRevD.94.055023}{{\em Phys.
  Rev.} {\bfseries D94} no.~5, (2016) 055023},
\href{http://arxiv.org/abs/1606.03107}{{\ttfamily arXiv:1606.03107 [hep-ph]}}.

\bibitem{Martini:2017ydu}
T.~Martini and P.~Uwer, ``{The Matrix Element Method at next-to-leading order
  QCD for hadronic collisions: Single top-quark production at the LHC as an
  example application},''
\href{http://arxiv.org/abs/1712.04527}{{\ttfamily arXiv:1712.04527 [hep-ph]}}.

\bibitem{Kraus:2019qoq}
M.~Kraus, T.~Martini, and P.~Uwer, ``{Predicting event weights at
  next-to-leading order QCD for jet events defined by $2\rightarrow 1$ jet
  algorithms},''
\href{http://arxiv.org/abs/1901.08008}{{\ttfamily arXiv:1901.08008 [hep-ph]}}.

\bibitem{Soper:2014rya}
D.~E. Soper and M.~Spannowsky, ``{Finding physics signals with event
  deconstruction},'' \href{http://dx.doi.org/10.1103/PhysRevD.89.094005}{{\em
  Phys. Rev.} {\bfseries D89} no.~9, (2014) 094005},
\href{http://arxiv.org/abs/1402.1189}{{\ttfamily arXiv:1402.1189 [hep-ph]}}.

\bibitem{Englert:2015dlp}
C.~Englert, O.~Mattelaer, and M.~Spannowsky, ``{Measuring the Higgs-bottom
  coupling in weak boson fusion},''
  \href{http://dx.doi.org/10.1016/j.physletb.2016.02.074}{{\em Phys. Lett.}
  {\bfseries B756} (2016) 103--108},
\href{http://arxiv.org/abs/1512.03429}{{\ttfamily arXiv:1512.03429 [hep-ph]}}.

\bibitem{Prestel:2019neg}
S.~Prestel and M.~Spannowsky, ``{HYTREES: Combining Matrix Elements and Parton
  Shower for Hypothesis Testing},''
\href{http://arxiv.org/abs/1901.11035}{{\ttfamily arXiv:1901.11035 [hep-ph]}}.

\bibitem{FerreiradeLima:2017iwx}
D.~E. Ferreira~de Lima, O.~Mattelaer, and M.~Spannowsky, ``{Searching for
  processes with invisible particles using a matrix element-based method},''
  \href{http://dx.doi.org/10.1016/j.physletb.2018.10.044}{{\em Phys. Lett.}
  {\bfseries B787} (2018) 100--104},
\href{http://arxiv.org/abs/1712.03266}{{\ttfamily arXiv:1712.03266 [hep-ph]}}.

\bibitem{Alwall:2014hca}
J.~Alwall, R.~Frederix, S.~Frixione, V.~Hirschi, F.~Maltoni, O.~Mattelaer,
  H.~S. Shao, T.~Stelzer, P.~Torrielli, and M.~Zaro, ``{The automated
  computation of tree-level and next-to-leading order differential cross
  sections, and their matching to parton shower simulations},''
  \href{http://dx.doi.org/10.1007/JHEP07(2014)079}{{\em JHEP} {\bfseries 07}
  (2014) 079},
\href{http://arxiv.org/abs/1405.0301}{{\ttfamily arXiv:1405.0301 [hep-ph]}}.

\bibitem{nofreelunch}
D.~H. {Wolpert} and W.~G. {Macready}, ``No free lunch theorems for
  optimization,'' \href{http://dx.doi.org/10.1109/4235.585893}{{\em IEEE
  Transactions on Evolutionary Computation} {\bfseries 1} no.~1, (April, 1997)
  67--82}.

\bibitem{nlopt}
{Steven G. Johnson}, ``{The NLopt nonlinear-optimization package}.''
  \url{http://github.com/stevengj/nlopt}.

\bibitem{10.1093/comjnl/20.4.367}
W.~L. Price, ``{A controlled random search procedure for global
  optimisation},'' \href{http://dx.doi.org/10.1093/comjnl/20.4.367}{{\em The
  Computer Journal} {\bfseries 20} no.~4, (01, 1977) 367--370}.
  \url{https://doi.org/10.1093/comjnl/20.4.367}.

\bibitem{Price1983}
W.~L. Price, ``Global optimization by controlled random search,''
  \href{http://dx.doi.org/10.1007/BF00933504}{{\em Journal of Optimization
  Theory and Applications} {\bfseries 40} no.~3, (Jul, 1983) 333--348}.
  \url{https://doi.org/10.1007/BF00933504}.

\bibitem{Kaelo2006}
P.~Kaelo and M.~M. Ali, ``Some variants of the controlled random search
  algorithm for global optimization,''
  \href{http://dx.doi.org/10.1007/s10957-006-9101-0}{{\em Journal of
  Optimization Theory and Applications} {\bfseries 130} no.~2, (Aug, 2006)
  253--264}. \url{https://doi.org/10.1007/s10957-006-9101-0}.

\bibitem{Jones1993}
D.~R. Jones, C.~D. Perttunen, and B.~E. Stuckman, ``Lipschitzian optimization
  without the lipschitz constant,''
  \href{http://dx.doi.org/10.1007/BF00941892}{{\em Journal of Optimization
  Theory and Applications} {\bfseries 79} no.~1, (Oct, 1993) 157--181}.
  \url{https://doi.org/10.1007/BF00941892}.

\bibitem{Gablonsky2001}
J.~Gablonsky and C.~Kelley, ``A locally-biased form of the direct algorithm,''
  \href{http://dx.doi.org/10.1023/A:1017930332101}{{\em Journal of Global
  Optimization} {\bfseries 21} no.~1, (Sep, 2001) 27--37}.
  \url{https://doi.org/10.1023/A:1017930332101}.

\bibitem{RinnooyKan1987.1}
A.~H.~G. Rinnooy~Kan and G.~T. Timmer, ``Stochastic global optimization methods
  part i: Clustering methods,''
  \href{http://dx.doi.org/10.1007/BF02592070}{{\em Mathematical Programming}
  {\bfseries 39} no.~1, (Sep, 1987) 27--56}.
  \url{https://doi.org/10.1007/BF02592070}.

\bibitem{RinnooyKan1987.2}
A.~H.~G. Rinnooy~Kan and G.~T. Timmer, ``Stochastic global optimization methods
  part ii: Multi level methods,''
  \href{http://dx.doi.org/10.1007/BF02592071}{{\em Mathematical Programming}
  {\bfseries 39} no.~1, (Sep, 1987) 57--78}.
  \url{https://doi.org/10.1007/BF02592071}.

\bibitem{Kucherenko2005}
S.~Kucherenko and Y.~Sytsko, ``Application of deterministic low-discrepancy
  sequences in global optimization,''
  \href{http://dx.doi.org/10.1007/s10589-005-4615-1}{{\em Computational
  Optimization and Applications} {\bfseries 30} no.~3, (Mar, 2005) 297--318}.
  \url{https://doi.org/10.1007/s10589-005-4615-1}.

\bibitem{873238}
T.~P. {Runarsson} and {Xin Yao}, ``Stochastic ranking for constrained
  evolutionary optimization,''
  \href{http://dx.doi.org/10.1109/4235.873238}{{\em IEEE Transactions on
  Evolutionary Computation} {\bfseries 4} no.~3, (Sep., 2000) 284--294}.

\bibitem{1424197}
T.~P. {Runarsson} and {Xin Yao}, ``Search biases in constrained evolutionary
  optimization,'' \href{http://dx.doi.org/10.1109/TSMCC.2004.841906}{{\em IEEE
  Transactions on Systems, Man, and Cybernetics, Part C (Applications and
  Reviews)} {\bfseries 35} no.~2, (May, 2005) 233--243}.

\bibitem{Beyer2002}
H.-G. Beyer and H.-P. Schwefel, ``Evolution strategies -- a comprehensive
  introduction,'' \href{http://dx.doi.org/10.1023/A:1015059928466}{{\em Natural
  Computing} {\bfseries 1} no.~1, (Mar, 2002) 3--52}.
  \url{https://doi.org/10.1023/A:1015059928466}.

\bibitem{5482078}
C.~H. {da Silva Santos}, M.~S. {Goncalves}, and H.~E. {Hernandez-Figueroa},
  ``Designing novel photonic devices by bio-inspired computing,''
  \href{http://dx.doi.org/10.1109/LPT.2010.2051222}{{\em IEEE Photonics
  Technology Letters} {\bfseries 22} no.~15, (Aug, 2010) 1177--1179}.

\bibitem{doi:10.1137/0728030}
A.~Conn, N.~Gould, and P.~Toint, ``A globally convergent augmented lagrangian
  algorithm for optimization with general constraints and simple bounds,''
  \href{http://dx.doi.org/10.1137/0728030}{{\em SIAM Journal on Numerical
  Analysis} {\bfseries 28} no.~2, (1991) 545--572},
  \href{http://arxiv.org/abs/https://doi.org/10.1137/0728030}{{\ttfamily
  https://doi.org/10.1137/0728030}}. \url{https://doi.org/10.1137/0728030}.

\bibitem{doi:10.1080/10556780701577730}
E.~Birgin and J.~Martínez, ``Improving ultimate convergence of an augmented
  lagrangian method,'' \href{http://dx.doi.org/10.1080/10556780701577730}{{\em
  Optimization Methods and Software} {\bfseries 23} no.~2, (2008) 177--195},
  \href{http://arxiv.org/abs/https://doi.org/10.1080/10556780701577730}{{\ttfamily
  https://doi.org/10.1080/10556780701577730}}.
  \url{https://doi.org/10.1080/10556780701577730}.

\bibitem{Powell2009TheBA}
M.~J. Powell, ``The bobyqa algorithm for bound constrained optimization without
  derivatives,'' Tech. Rep. NA2009/06, Department of Applied Mathematics and
  Theoretical Physics, Cambridge, 2009.

\bibitem{Powell1994}
M.~J.~D. Powell, {\em A Direct Search Optimization Method That Models the
  Objective and Constraint Functions by Linear Interpolation},
  \href{http://dx.doi.org/10.1007/978-94-015-8330-5_4}{pp.~51--67}.
\newblock Springer Netherlands, Dordrecht, 1994.
\newblock \url{https://doi.org/10.1007/978-94-015-8330-5_4}.

\bibitem{10.1093/comjnl/7.4.308}
J.~A. Nelder and R.~Mead, ``{A Simplex Method for Function Minimization},''
  \href{http://dx.doi.org/10.1093/comjnl/7.4.308}{{\em The Computer Journal}
  {\bfseries 7} no.~4, (01, 1965) 308--313}.
  \url{https://doi.org/10.1093/comjnl/7.4.308}.

\bibitem{10.1093/comjnl/8.1.42}
M.~J. Box, ``{A New Method of Constrained Optimization and a Comparison With
  Other Methods},'' \href{http://dx.doi.org/10.1093/comjnl/8.1.42}{{\em The
  Computer Journal} {\bfseries 8} no.~1, (04, 1965) 42--52}.
  \url{https://doi.org/10.1093/comjnl/8.1.42}.

\bibitem{Powell2006}
M.~J.~D. Powell, {\em The NEWUOA software for unconstrained optimization
  without derivatives},
  \href{http://dx.doi.org/10.1007/0-387-30065-1_16}{pp.~255--297}.
\newblock Springer US, Boston, MA, 2006.
\newblock \url{https://doi.org/10.1007/0-387-30065-1_16}.

\bibitem{praxis}
R.~Brent, {\em Algorithms for Minimization without Derivatives}.
\newblock Prentice-Hall, 1972.

\bibitem{Rowan90functionalstability}
T.~H. Rowan, ``Functional stability analysis of numerical algorithms,'' tech.
  rep., 1990.

\bibitem{Aaboud:2017rss}
{\bfseries ATLAS} Collaboration, M.~Aaboud {\em et~al.}, ``{Search for the
  standard model Higgs boson produced in association with top quarks and
  decaying into a $b\bar{b}$ pair in $pp$ collisions at $\sqrt{s}$ = 13 TeV
  with the ATLAS detector},''
  \href{http://dx.doi.org/10.1103/PhysRevD.97.072016}{{\em Phys. Rev.}
  {\bfseries D97} no.~7, (2018) 072016},
\href{http://arxiv.org/abs/1712.08895}{{\ttfamily arXiv:1712.08895 [hep-ex]}}.

\bibitem{Sirunyan:2018mvw}
{\bfseries CMS} Collaboration, A.~M. Sirunyan {\em et~al.}, ``{Search for $
  \mathrm{t}\overline{\mathrm{t}}\mathrm{H} $ production in the $ \mathrm{H}\to
  \mathrm{b}\overline{\mathrm{b}} $ decay channel with leptonic $
  \mathrm{t}\overline{\mathrm{t}} $ decays in proton-proton collisions at $
  \sqrt{s}=13 $ TeV},'' \href{http://dx.doi.org/10.1007/JHEP03(2019)026}{{\em
  JHEP} {\bfseries 03} (2019) 026},
\href{http://arxiv.org/abs/1804.03682}{{\ttfamily arXiv:1804.03682 [hep-ex]}}.

\bibitem{Ball:2013hta}
{\bfseries NNPDF} Collaboration, R.~D. Ball, V.~Bertone, S.~Carrazza,
  L.~Del~Debbio, S.~Forte, A.~Guffanti, N.~P. Hartland, and J.~Rojo, ``{Parton
  distributions with QED corrections},''
  \href{http://dx.doi.org/10.1016/j.nuclphysb.2013.10.010}{{\em Nucl. Phys.}
  {\bfseries B877} (2013) 290--320},
\href{http://arxiv.org/abs/1308.0598}{{\ttfamily arXiv:1308.0598 [hep-ph]}}.

\bibitem{Buckley:2014ana}
A.~Buckley, J.~Ferrando, S.~Lloyd, K.~Nordstroem, B.~Page, M.~Ruefenacht,
  M.~Schoenherr, and G.~Watt, ``{LHAPDF6: parton density access in the LHC
  precision era},''
  \href{http://dx.doi.org/10.1140/epjc/s10052-015-3318-8}{{\em Eur. Phys. J.}
  {\bfseries C75} (2015) 132},
\href{http://arxiv.org/abs/1412.7420}{{\ttfamily arXiv:1412.7420 [hep-ph]}}.

\bibitem{Aaboud:2018zhk}
{\bfseries ATLAS} Collaboration, M.~Aaboud {\em et~al.}, ``{Observation of $H
  \rightarrow b\bar{b}$ decays and $VH$ production with the ATLAS detector},''
  \href{http://dx.doi.org/10.1016/j.physletb.2018.09.013}{{\em Phys. Lett.}
  {\bfseries B786} (2018) 59--86},
\href{http://arxiv.org/abs/1808.08238}{{\ttfamily arXiv:1808.08238 [hep-ex]}}.

\bibitem{deFlorian:2016spz}
{\bfseries LHC Higgs Cross Section Working Group} Collaboration, D.~de~Florian
  {\em et~al.}, ``{Handbook of LHC Higgs Cross Sections: 4. Deciphering the
  Nature of the Higgs Sector},''
\href{http://arxiv.org/abs/1610.07922}{{\ttfamily arXiv:1610.07922 [hep-ph]}}.

\end{thebibliography}\endgroup

\end{document}